Plasma treatment as an unconventional molecular magnet engineering method

D. Czernia[*,1], P. Konieczny[1], M. Perzanowski[1], and D. Pinkowicz[2]

[1]Institute of Nuclear Physics Polish Academy of Sciences, Radzikowskiego 152, 31-342 Kraków, Poland

[2]Department of Chemistry, Jagiellonian University, Gronostajowa 2, 30-387 Kraków, Poland

Molecular magnetism aims to design materials with unique properties at the molecular level, focusing on the systematic synthesis of new chemical compounds. In this paper, we propose an alternative route to engineer molecular magnetic materials through plasma irradiation. Our research indicates that the long-range magnetic order temperature in the three-dimensional $\{[Mn^{II}(H_2O)_2]_2[Nb^{IV}(CN)_8]\cdot 4H_2O\}_n$ molecular ferrimagnet increases by 20 K after plasma treatment. The core structure of the compound does not reveal significant changes after plasma processing, as confirmed by the X-ray powder diffraction analysis. The observed results are attributed to the release of crystallized water molecules. The described procedure can serve as a viable approach to altering the magnetic properties of the molecular systems.

1. Introduction

Molecular magnetic materials have emerged at the confluence of chemistry, physics, and materials science, holding promise for a wide range of applications from spintronics or qubits designs [1] to magnetic refrigeration [2] or optically active sensors and switches [3,4]. The pursuit of molecular magnets stems from the desire to engineer materials at the molecular level, enabling precise control over their magnetic behavior, usually done by selecting proper synthesis components and procedures. The alternative post-synthesis approaches to exploit the potential of molecular magnetic materials include incorporating them into thin films [5] or influencing their properties with ion irradiation [6].

Plasma modification provides another way to manipulate magnetism through surface modification [7,8] or by introducing specific chemical changes and structural alternations. This can lead to changes in the ferromagnetic interactions [9], magnetic hysteresis loops [10], and blocking temperature [11]. Plasma treatment has been applied in the functionalization [12], polymerization [13], and surface activation and grafting of polymers [14], which share similarities with molecular magnets regarding chemical bonding and composition. Notably, there have been no reports on the effect of plasma irradiation on molecular magnetic materials.

Here, we present the first case of combining plasma modification technique with molecular magnetism in the octacyanidometalate $\{[Mn^{II}(H_2O)_2]_2[Nb^{IV}(CN)_8]\cdot 4H_2O\}_n$ ferrimagnet, representing the well-studied isostructural family of compounds [15–21]. This coordination polymer comprises a three-dimensional network with channels occupied by water molecules coordinated to $Mn^{II}$ and crystallized ones. It is a soft magnet with the long-range magnetic order (LRMO) transition of $T_C = 49$ K. This report demonstrates that it is possible to obtain another magnetic phase with a 20 K higher $T_C$ using air-based plasma with little interference in the structure of the chemical compound, i.e., in the original three-dimensional network of magnetic ions linked through cyanide ligands, as observed in the X-ray powder diffraction patterns. This may indicate the loss of water molecules that alter the distance between $Mn^{II}$ and $Nb^{IV}$ and modify antiferromagnetic exchange interactions.

---

[*] Corresponding author: dominik.czernia@ifj.edu.pl

## 2. Materials and methods

### 2.1. Sample preparation

The studied $\{[Mn^{II}(H_2O)_2]_2[Nb^{IV}(CN)_8]\cdot 4H_2O\}_n$ coordination polymer (hereafter denoted as **NbMn$_2$**) was synthesized using $MnCl_2\cdot 4H_2O$ and $K_4[Nb(CN)_8]\cdot 2H_2O$ as per previously published procedures [15]. This led to the formation of dark red crystalline specimens. The powder was ground manually in a mortar to decrease the crystallite size and scattered on a 1×1 cm$^2$ strip of Scotch tape, forming a thin sample layer with a mass of about 0.2 mg, guaranteeing a high surface-to-volume ratio for further processing.

The samples were exposed to an air plasma generated by the Harrick Plasma Cleaner in high-power mode (18 W) for 2 (**NbMn$_2$-2**), 10 (**NbMn$_2$-10**), and 15 (**NbMn$_2$-15**) minutes. The color of the **NbMn$_2$-10** and **NbMn$_2$-15** turned dark brown. Subsequently, an extra layer of Scotch tape was placed on the powder-taped samples, which were then folded and encapsulated inside a gelatin capsule with the high vacuum Apiezon M grease for protection against atmospheric air. The infrared thermometer recorded a temperature inside the plasma cleaner not exceeding 75 °C.

### 2.2. Characterization methods

X-ray powder diffraction (XRPD) patterns were obtained by the PANalytical X'Pert Pro instrument with a copper X-ray tube source (Cu K$_{\alpha 1}$, 1.541 Å) in the θ-2θ geometry operating at 40 kV and 30 mA. Data were recorded at room temperature between 5° and 45° of 2θ. Background correction was applied, and the Scotch tape influence and the anode's characteristic X-ray K$_{\alpha 2}$ were removed.

The magnetic properties were measured using the Quantum Design SQUID MPMS-XL magnetometer. The isothermal magnetization $M(\mu_0 H)$ was obtained with an applied magnetic field of $\mu_0 H = [0, 7\text{ T}]$. Static magnetic susceptibility measurements were carried out while cooling in the temperature range of $T = 2.0\text{-}300$ K in the applied field of $\mu_0 H_{dc} = 500$ G. The temperature-independent diamagnetic contribution to susceptibility from the sample and its protection was subtracted by fitting the experimental data to the high-temperature range of measured magnetic susceptibility.

## 3. Results and discussion

### 3.1 X-ray powder diffraction (XRPD)

Detailed information on the **NbMn$_2$** crystallographic structure can be found in the reference article [16]. **NbMn$_2$** has a three-dimensional network and crystallizes in the tetragonal space group *I4/m* with $a = b = 12.080(2)$ Å and $c = 13.375(4)$ Å. The structure consists of alternately linked $Nb^{IV}$ and two $Mn^{II}$ ions through the cyanide ligands (CN$^-$). The $Nb^{IV}$ ion is coordinated by eight carbon atoms from CN$^-$, forming an approximately square antiprismatic coordination. Meanwhile, $Mn^{II}$ is coordinated by two $H_2O$ and four CN$^-$ (by nitrogen) molecules, resulting in a slightly distorted octahedron (Fig. 1). The shortest distance between the $Nb^{IV}$ and $Mn^{II}$ ions is either 5.539 Å or 5.482 Å. Moreover, the network contains four crystallized water molecules.

X-ray powder diffraction patterns acquired for the reference and **NbMn$_2$-15** reveal the presence of the same peaks in analog positions (Fig. 2). Thus, no significant structural variations in **NbMn₂** occurred throughout the plasma treatment. However, the peaks of the **NbMn₂-15** are slightly broader and shifted towards higher 2θ, by 0.17° on average. The peak shift indicates crystal lattice shrinkage, while their greater half-widths may result from increased microstrains in the crystal lattice or crystallite size reduction.

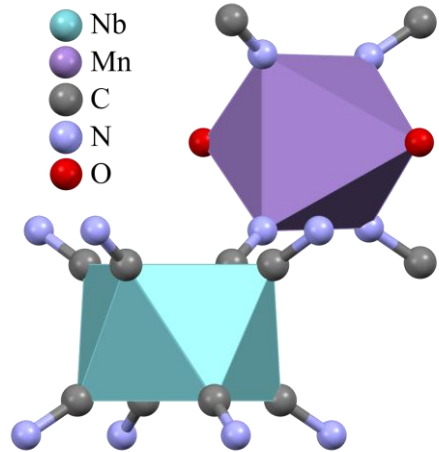

Fig. 1. Representation of the Nb$^{IV}$ approximately square antiprism (light blue) and Mn$^{II}$ distorted octahedron (violet) coordinations. The hydrogen atoms are omitted. The complete crystal structure can be found in reference [16].

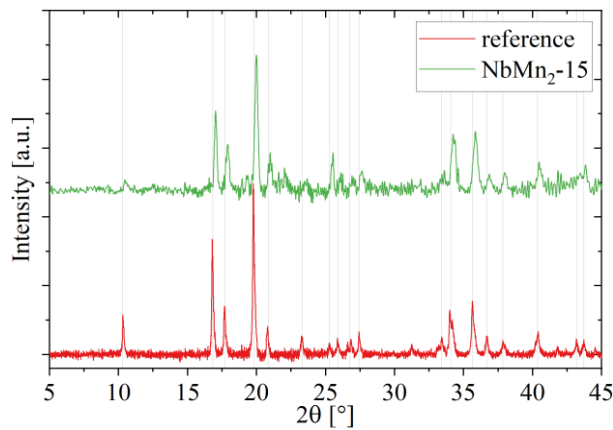

Fig. 2. X-ray powder diffraction patterns for the samples on Scotch tape before (red line; bottom) and after 15 minutes of plasma treatment, **NbMn₂-15** (green line; top), measured in the 2θ range of 5-45°. Gray lines indicate the maxima for the reference sample.

3.2 Magnetic properties

The magnetic properties of **NbMn₂** have been investigated in previous studies [16], revealing the LRMO below $T_C$ = 49 K and without opening the isothermal magnetization curve down to $T$ = 2.0 K. The magnetization saturation reaches 9 $\mu_B \cdot mol^{-1}$, which corresponds to the anticipated value for the antiparallel configuration of Nb$^{IV}$ ($S$ = 1/2; $g$ = 2.0) and Mn$^{II}$ ($S$ = 5/2; $g$ = 2.0) ions within the NbMn₂ unit. The mean-field approximation predicts the antiferromagnetic superexchange coupling constant of $J_{NbMn}$ = -15.44(7) K between Nb$^{IV}$ and Mn$^{II}$ ions, as well as an antiferromagnetic interaction within the Mn sublattice that is at least an order of magnitude smaller [15].

The isothermal magnetization field dependence for studied samples measured at $T = 2.0$ K from $\mu_0H = 0$ to 7 T is shown in Fig. 3. The magnetization curves all overlap above approximately $\mu_0H = 2$ T at the expected level for isotropic ions of 9 $\mu_B \cdot$mol$^{-1}$. However, the magnetization saturates slightly slower for the **NbMn$_2$-10** and visibly slower for the **NbMn$_2$-15**, which indicates a weak modification of the magnetization process caused by incorporating plasma-induced defects leading to the pinning of domain walls.

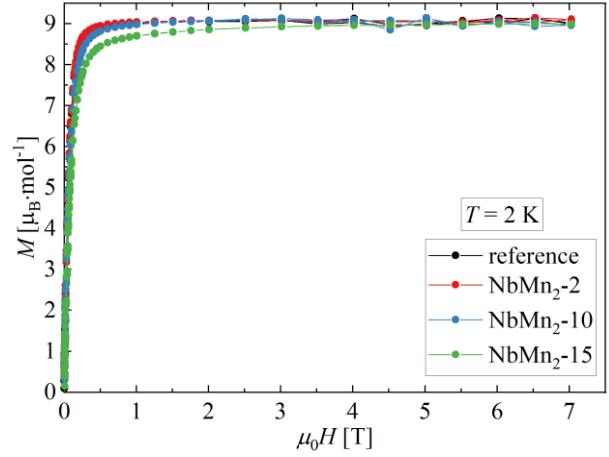

Fig. 3. The isothermal magnetization measured at $T = 2.0$ K in the applied magnetic fields $\mu_0H = 0$-7 T of the **NbMn$_2$** samples.

More prominent changes can be seen in the magnetic susceptibility $\chi$ measured in $\mu_0H = 500$ G from 300 K to 2 K, as illustrated in Fig. 4. The results are presented in the form of $\chi T$ product showing the maximum at $T_{max} \approx 40$ K for the reference, **NbMn$_2$-2** and **NbMn$_2$-10**. For the **NbMn$_2$-15**, it was $T_{max} \approx 56$ K. The gradual shift in $T_{max}$ can be observed in $\chi T$ for **NbMn$_2$-10** as an obvious two-peak susceptibility composition, indicating the emergence of a second magnetic phase with higher $T_C$.

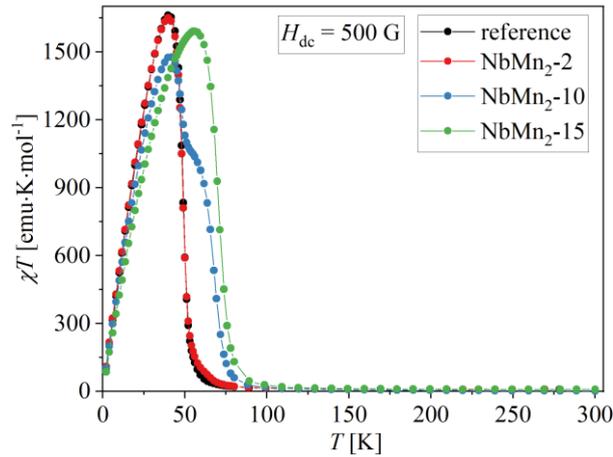

Fig. 4. Temperature dependence of the $\chi T$ product for **NbMn$_2$** measured in the magnetic field of $\mu_0H_{dc} = 500$ G. Lines are a guide for the eye.

The $T_C$ value was estimated from the minima of the $\chi T(T)$ first derivative: 49 K each for the reference and **NbMn$_2$-2**; two minima: 48 K and 67.5 K for **NbMn$_2$-10**; and 69.5 K for **NbMn$_2$-15**. The observed outcomes can be explained by the water molecules that escape from the three-dimensional network channels as a consequence of plasma-lattice interaction, similar to the desorption studies in other Nb/Mn-based octacyanidometalates [22–25]. Removing water molecules leads to the crystal lattice shrinkage, as seen in XRPD, thus decreasing the $Mn^{II}$-$Nb^{IV}$ distance and enhancing exchange coupling interactions.

4. Conclusions

Our study presents the first example of applying plasma modification to molecular magnetic materials to tailor their selected properties. We investigated the impact of plasma treatment on the three-dimensional **NbMn$_2$** molecular ferrimagnet. A second magnetic phase emerges after plasma treatment with a critical temperature $T_C$ increased by 20.5 K from its initial value of $T_C = 49$ K, reaching $T_C = 69.5$ K. In contrast, the magnetization process $T = 2.0$ K undergoes only slight modification. The measurements of XRPD indicate only subtle modifications of the crystal structure. Additionally, the process of creating this new phase is gradual and can be regulated through plasma irradiation time.

Further studies are needed to explore the nature of the modification in magnetic properties. High temperature and ultraviolet light generated in plasma, as well as desorption of crystallized and coordinated water, are potential factors contributing to the observed results.

Acknowledgments

This work was supported by the Polish Minister of Education and Science (G. No.: DI2017 006047).